\documentclass[journal, onecolumn, 12pt]{IEEEtran}

\usepackage{amsmath}
\usepackage{subfigure}
\usepackage{url}
\usepackage{ifpdf}
\usepackage{graphicx}
\usepackage{mathtools}
\usepackage{cite}
\usepackage{setspace}
\DeclarePairedDelimiter\floor{\lfloor}{\rfloor}

\begin{document}

\title{On the Performance of Network Coding and Forwarding Schemes with Different Degrees of Redundancy for Wireless Mesh Networks}
\author{Manolis Ploumidis, Nikolaos Pappas, Vasilios A. Siris, Apostolos Traganitis
\thanks{M. Ploumidis was supported by HERACLEITUS II -
University of Crete, NSRF (ESPA) (2007-2013) and is co-funded by the
European Union and national resources.}
\thanks{M.~Ploumidis is with the
Institute of Computer Science, Foundation for Research and Technology - Hellas (FORTH)
and Computer Science Department, University of Crete, Greece email:ploumid@ics.forth.gr}
\thanks{N.~Pappas is with the Sup\'{e}lec, Department of Telecommunications, Gif-sur-Yvette, France email:nikolaos.pappas@supelec.fr}
\thanks{V.~A.~Siris is with the Athens University of Economics and Business, Greece, email: vsiris@aueb.gr}
\thanks{A.~Traganitis is with the with the
Institute of Computer Science, Foundation for Research and Technology - Hellas (FORTH)
and Computer Science Department, University of Crete, email: tragani@ics.forth.gr.}}

 \maketitle

\begin{abstract}
This study explores the throughput and delay that can be achieved by various forwarding schemes employing multiple paths and different degrees of redundancy focusing on linear network coding.
The key contribution of the study is an analytical framework for modeling the throughput and delay for various schemes considering wireless mesh networks where, unicast traffic is forwarded and hop-by-hop retransmissions
are employed for achieving reliability.
The analytical framework is generalized for an arbitrary number of paths and hops per path.
Another key contribution of the study is the evaluation and extension of the numerical results drawn from the analysis through NS-2 simulations.
Our results show that in scenarios with significant interference the best throughput-delay tradeoff is achieved by single path forwarding.
Moreover, when significant interference is present and network coding employs the larger packet generation size it experiences higher delay than all other schemes due to the inter-arrival times
aggregating over all coded packets required to decode a packet generation.

\end{abstract}


\section{\uppercase{Introduction}}
\label{sec:introd}

Utilization of multiple paths in parallel in wireless networks can provide a wide range of benefits in terms of, throughput \cite{6133896}, delay \cite{mp_route_wim}, load balancing \cite{5198989} e.t.c.
Jointly employing multiple paths and redundancy has been adopted by various schemes aimed at increasing reliability \cite{Oh:2009:RMR:1702135.1702167}.

The idea of using redundancy is central in channel coding theory.
The work in~\cite{b:divcod} uses path diversity for fast recovery from link outages.
The work in~\cite{b:yeungerrcorr} introduces error correcting network coding as a generalization of classical error correcting codes.
Network coding is a generalization of the traditional store and forward technique.
The core notion of network coding introduced in~\cite{b:yeung} is to allow and encourage mixing of data at intermediate network nodes.
The work in~\cite{b:lun} studies the coding delay in packet networks that support network coding.
The authors in~\cite{b:designeffrobnet} propose efficient algorithms for the construction of robust network codes for multicast connections.
The work in~\cite{b:robnetcodpathfailure} presents an approach for designing network codes by considering path failures in the network instead of edge failures.
There is a lot of work concerning opportunistic routing in wireless mesh networks, with or without network coding.
COPE~\cite{b:cope}, MORE~\cite{b:more} and MC$^{2}$~\cite{b:mc2} investigate network coding with opportunistic routing in wireless networks with broadcast transmissions, focusing exclusively on the throughput improvements.
ExOR~\cite{b:exor} and ROMER~\cite{b:romer} investigate opportunistic routing in broadcast wireless networks without network coding.
Moreover, these works also focus on the throughput improvements, except~\cite{b:romer} which also considers the packet delivery ratio.
The work of~\cite{b:tsirig1} considers diversity coding, and investigates the allocation of data to multiple paths that maximizes the probability of successful reception.
The work of~\cite{b:tsirig2} extends the previous work, in the case where the failure probabilities are different for different paths, and when the paths are not necessarily independent.
In \cite{Oh:2009:RMR:1702135.1702167} the authors suggest and evaluate through simulations an adaptive multipath routing protocol that switches between single path, multipath with network coding and multipath routing that
replicates packets on all paths based on the observed channel loss conditions.
In \cite{netcod_des:2013} the authors discuss several issues that affect the performance in terms of computational complexity of practical network coding implementations
including network coding parameters, such as, generation and field size and also platform dependent and protocol related issues.
The work in \cite{6134700} explores the rate-delay tradeoff for a multipath transmission scheme employing multiple disjoint paths and network coding at the source seeking for the optimal rate while also
satisfying end-to-end delay based QoS requirements.
CoMP suggested in \cite{5502204} is a multipath online network coding scheme that is aimed at improving the performance of TCP sessions in multihop wireless mesh networks.
The rate at which linear independent combinations are injected in the network depends on estimates of link loss rates.

Most of the theoretical results in network coding consider multicast traffic but the vast majority of Internet traffic is unicast.
Applying network coding in wireless environments has to address multiple unicast flows, if it has any chance of being used.
Especially for the case of multicast traffic where all receivers are interested for all packets, intermediate nodes can encode any packets together, without worrying about decoding which will be performed eventually at the destinations.

In this study, we consider wireless mesh networks where multiple paths are available between a source and a destination node.
Source and destination are equipped with multiple interfaces and the type of traffic forwarded is unicast.
For achieving reliability, hop-by-hop retransmissions are assumed.
In short, the forwarding schemes explored are: \textit{single path} that employs zero redundancy and one path, \textit{multipath} that employs multiple paths and zero redundancy, \textit{multicopy} that replicates each packet on every path, and \textit{network coding-based forwarding}.

The key contribution of the study is an analytical framework for modeling the throughput and delay of the aforementioned forwarding schemes.
In the first part of the analysis we express the throughput and delay for all forwarding schemes for a simple topology considering an erasure wireless channel where link error probability for each link
is captured through the SINR model and demonstrate the complexity of generalizing for arbitrary topologies.
In the second part of the analysis, the framework is generalized for an arbitrary number of paths and hops per paths where link error probabilities are expressed through the SNR model.
The second key contribution of this study is the validation and extension of the numerical results drawn from the analytical framework through NS-2 simulation for realistic wireless settings.

The simulation results show that in scenarios with significant interference best delay and throughput is achieved by forwarding schemes that moderate the parallel use of paths with the
best throughput delay trade-off achieved by single path forwarding.
In the presence of high interference our analytical framework underestimates the rank of single path forwarding both in terms of delay and throughput.
Moreover, when significant interference is present and network coding employs a larger packet generation size it experiences higher delay than all other schemes due to the inter-arrival times
aggregating over all coded packets required to decode a packet generation.
Finally in scenarios with lower interference the suggested framework overestimates only the rank of network coding in terms of delay.

\section{\uppercase{System Model}}
\label{sec:system_model}

We assume a wireless acyclic network of a single source sending unicast traffic to a single destination node through multiple paths that consist of lossy links.
The paths available between the source and the destination can be either node-disjoint or share common nodes and are assumed to be given by some multi-path routing protocol \cite{multipath_olsr}.
Moreover, source routing is assumed ensuring that packets of the same flow will be forwarded to the destination through the same path.
For achieving reliability, hop-by-hop retransmissions are assumed.
When an error occurs at the transmission of a packet between two nodes for example node $i$ and $i + 1$, node $i$ retransmits the packet to $i + 1$.
Acknowledgements for successfully received packets are assumed to be instantaneous and error free.
Further on, we assume that there is no congestion, hence no queuing delay, at the intermediate nodes.
For the case of network coding, when the network has more than one hop, the inner nodes can decode the information and then re-encode it.

In this work we model delay and throughput achieved for the following schemes:
\begin{itemize}
\item Single path forwarding (SP) - Only one path is employed in order to forward traffic to the destination
\item Multicopy (MC) - Each packet is replicated on all available
paths employing in this way the maximum possible redundancy
\item Multipath (MP) - Each packet is assigned on a specific path with different packets of a flow being assigned on different paths. It employs zero redundancy.
\item Network-coding based forwarding (NC) - Data packets are grouped in sets of size $k$ constituting different \textit{packet generations}.
Packets of each packet generation are coded together through linear network coding resulting in $m={2}^{k}-1$ linearly independent combinations excluding the combination that contains only zero values.
Each such linear independent combination constitutes a coded packet that is assigned on a specific path.
A packet generation can be decoded and the original data can be extracted if $k$ or more coded packets are received.
All coded packets are forwarded in parallel.
\end{itemize}

In this work we disregard the queuing delay at the sender, the encoding and decoding delays, and the ACK transmission delays.
In the first part of the analysis presented in section \ref{sec:analysis_sinr_model} we consider a wireless erasure channel where link error probability for each link is captured through the SINR model.

In the wireless environment, a packet can be decoded correctly by the receiver if the received $SINR$ exceeds a certain threshold.
More precisely, suppose that we are given a set $T$ of nodes transmitting in the same time slot.
Let  $P_{rx}(i,j)$ be the signal power received from node $i$ at node $j$.
Let $SINR(i,j)$ be expressed using (\ref{eq:sinr_thres}).
\begin{equation}
\label{eq:sinr_thres}
SINR(i,j)=\frac{P_{rx}(i,j)}{\eta_{j}+\sum_{k\in T\backslash\left\{i\right\}} {P_{rx}(k,j)}}.
\end{equation}
In the above equation $\eta_{j}$ denotes the receiver noise power at $j$. We assume that a packet transmitted by $i$ is successfully received by $j$
if and only if $SINR(i,j)\geq \gamma_{j}$, where $\gamma_{j}$ is a threshold characteristic of node $j$. The wireless channel is subject to fading;
let $P_{tx}(i)$ be the transmitting power of node $i$ and $r(i,j)$ be the distance between $i$ and $j$. The power received by $j$ when $i$
transmits is $P_{rx}(i,j)=A(i,j)g(i,j)$ where $A(i,j)$ is a random variable representing channel fading. Under Rayleigh fading, it is
known~\cite{b:Tse} that $A(i,j)$ is exponentially distributed. The received power factor $g(i,j)$ is given by $g(i,j)=P_{tx}(i)(r(i,j))^{-\alpha}$
where $\alpha$ is the path loss exponent with typical values between $2$ and $4$. The success probability of link $(i,j)$ when the transmitting nodes
are in $T$ is given by

\begin{equation}
\label{eq:succprob}
p_{i/T}^{j}=\exp\left(-\frac{\gamma_{j}\eta_{j}}{v(i,j)g(i,j)}\right) \prod_{k\in T\backslash \left\{i,j\right\}}{\left(1+\gamma_{j}\frac{v(k,j)g(k,j)}{v(i,j)g(i,j)}\right)}^{-1},
\end{equation}

where $v(i,j)$ is the parameter of the Rayleigh random variable for fading. The analytical derivation for this success probability
which captures the effect of interference on link $(i,j)$ from transmissions of nodes in set $T$, can be found in~\cite{b:Nguyen}.
In the second part of the analysis presented in section \ref{sec:analysis_snr_model} we relax the assumption concerning the wireless channel capture link error probability through the SNR model.

\section{\uppercase{Analysis}}
\label{sec:analysis}

Before proceeding with the analysis for the average delay and throughput the following definitions are required:
For the case of single path, multipath and multicopy, packet delay \textit{D} is the delay for transmitting a packet from the source to the destination when the packet is at the head of the transmission queue at the source node.
We also assume that the transmission of one packet requires one time slot.
If a packet is not correctly received by the the destination or intermediate nodes it is retransmitted on a hop-by-hop manner.
For the case of network coding based schemes, assuming a packet generation of size \textit{k}, delay is estimated as the average delay to receive at least \textit{k} coded packets.

\subsection{Link Error Probabilities Based on the SINR Model}
\label{sec:analysis_sinr_model}

In this section throughput and delay is expressed for all aforementioned forwarding schemes for a network consisting of three single hop paths where link error probability is determined
based on the SINR model presented in section \ref{sec:system_model}.

- Single or Best Path: the link $j$ (path) with the lowest link error probability is selected to forward traffic to the destination provided by:
\begin{equation}
j=\arg\min_{i} e_{i/i},\text{ }i=1,2,3,
\end{equation}
the delay is given by
\begin{equation}
D_{sp}=\frac{1}{1-e_{j/j}},
\end{equation}
where, $e_{i/i} = 1-p_{i/i}$ with $p_{i/i}$ provided by equation \ref{eq:succprob}. The throughput is given by $Thr_{sp}=1/D_{sp}$.

- Multipath: The packets are transmitted in parallel through all available paths.The delay for multipath is $D_{mp} = D_{mp,3} / 3$, where

\begin{equation}
\begin{aligned}
D_{mp,3} = (1-e_{1/1,2,3})(1-e_{2/1,2,3})(1-e_{3/1,2,3}) + (1-e_{1/1,2,3})(1-e_{2/1,2,3})e_{3/1,2,3} D_{mp}^{3} + \\
+(1-e_{1/1,2,3})(1-e_{3/1,2,3})e_{2/1,2,3} D_{mp}^{2} +(1-e_{2/1,2,3})(1-e_{3/1,2,3})e_{1/1,2,3} D_{mp}^{1}+ \\
+(1-e_{1/1,2,3})e_{2/1,2,3}e_{3/1,2,3} D_{mp}^{2,3}
+(1-e_{2/1,2,3})e_{1/1,2,3}e_{3/1,2,3} D_{mp}^{1,3}+\\
+(1-e_{3/1,2,3})e_{1/1,2,3}e_{2/1,2,3} D_{mp}^{1,2}
+e_{1/1,2,3}e_{2/1,2,3}e_{3/1,2,3}(1+D_{mp,3}).
\end{aligned}
\end{equation}

$D_{mp,3}$ is the average delay to receive the three packets from the three paths. $D_{mp}^{i}$ for $i=1,2,3$, is the delay to receive the packet from $i$-th path and $D_{mp}^{i} = \frac{1}{1-e_{i/i}}$.
When only the packet from the first path is received, then the delay to receive the rest two packets from the second and the third path is $D_{mp}^{2,3}$, and is given by
\begin{equation}
D_{mp}^{2,3} =   (1-e_{1/1,2})(1-e_{2/1,2}) + (1-e_{1/1,2})e_{2/1,2} D_{mp}^{2} + (1-e_{2/1,2})e_{1/1,2} D_{mp}^{1}+ e_{1/1,2}e_{2/1,2}(1+D_{mp}^{2,3}),
\end{equation}
$D_{mp}^{1,2}$ and $D_{mp}^{1,3}$ can be calculated in the same way and thus the corresponding calculation is omitted.

The achieved throughput is $Thr_{mp}=3/D_{mp}$.

- Multicopy (MC): The delay is
\begin{equation}
\begin{aligned}
D_{mc} = (1-e_{1/1,2,3})(1-e_{2/1,2,3})(1-e_{3/1,2,3}) + (1-e_{1/1,2,3})(1-e_{2/1,2,3})e_{3/1,2,3} + \\
+(1-e_{1/1,2,3})(1-e_{3/1,2,3})e_{2/1,2,3} +(1-e_{2/1,2,3})(1-e_{3/1,2,3})e_{1/1,2,3}+ \\
+(1-e_{1/1,2,3})e_{2/1,2,3}e_{3/1,2,3}
+(1-e_{2/1,2,3})e_{1/1,2,3}e_{3/1,2,3}+\\
+(1-e_{3/1,2,3})e_{1/1,2,3}e_{2/1,2,3}
+e_{1/1,2,3}e_{2/1,2,3}e_{3/1,2,3}(1+D_{mc}).
\end{aligned}
\end{equation}
The throughput is given by $Thr_{mc}=1/D_{mc}$.

- Multipath with Network Coding:
Assuming a packet generation of size two, applying linear network coding on two data packets results in three coded packets and a fourth one containing only zero values.
In order to successfully decode a packet generation thus, we need to receive two or three coded packets.
If only one coded packet is received through path \textit{i} the receiver will wait for the other paths to accomplish
a successful coded packet delivery. Thus the delay is

\begin{equation}
\begin{aligned}
D_{nc} = (1-e_{1/1,2,3})(1-e_{2/1,2,3})(1-e_{3/1,2,3}) + (1-e_{1/1,2,3})(1-e_{2/1,2,3})e_{3/1,2,3} + \\
+(1-e_{1/1,2,3})(1-e_{3/1,2,3})e_{2/1,2,3} +(1-e_{2/1,2,3})(1-e_{3/1,2,3})e_{1/1,2,3}+ \\
+(1-e_{1/1,2,3})e_{2/1,2,3}e_{3/1,2,3}(1+D_{nc}^{1})
+(1-e_{2/1,2,3})e_{1/1,2,3}e_{3/1,2,3}(1+D_{nc}^{2})+ \\
+(1-e_{3/1,2,3})e_{1/1,2,3}e_{2/1,2,3}(1+D_{nc}^{3})
+e_{1/1,2,3}e_{2/1,2,3}e_{3/1,2,3}(1+D_{nc}).
\end{aligned}
\end{equation}

In the previous equation, $D_{nc}^{1}$ denotes the delay required to receive at least one more coded packet given that the destination has already received one from the first path and is given by the following expression:

\begin{equation}
\begin{aligned}
D_{nc}^{1} = (1-e_{2/2,3})(1-e_{3/2,3})+ (1-e_{2/2,3})e_{3/2,3}
+e_{2/2,3}(1-e_{3/2,3})+ e_{2/2,3}e_{3/2,3} (1+D_{nc}^1).
\end{aligned}
\end{equation}

$D_{nc}^{2}$ and $D_{nc}^{3}$ can be calculated in the same way and thus the corresponding calculation is omitted.

The throughput for network coding is $Thr_{nc}=2/D_{nc}$.

The previous analysis reveals that analytically expressing the throughput and delay achieved by the aforementioned schemes requires exhaustive enumeration of all possible sets of interfering transmitters.
For larger topologies consisting of multiple multi-hop paths where intra-path interference may also be present such an approach would be computationally intractable.
This process is further complicated if transmission probabilities are adopted for each source and relay node.

\subsection{Link Error Probabilities based on the SNR Model}
\label{sec:analysis_snr_model}
In this section we express the delay and throughput of all the aforementioned schemes
considering different network settings based on the following parameters:
\begin{itemize}
\item Symmetric or not symmetric links in terms of error probability
\item Paths being either disjoint or not
\item End-to-end or hop-by-hop coding process for network coding based schemes.
\end{itemize}

\subsubsection{Node-disjoint Paths, End-to-End Coding}
\label{sec:analysis_node_disjoint_paths}

\begin{figure}
\begin{center}
\includegraphics[scale=0.5]{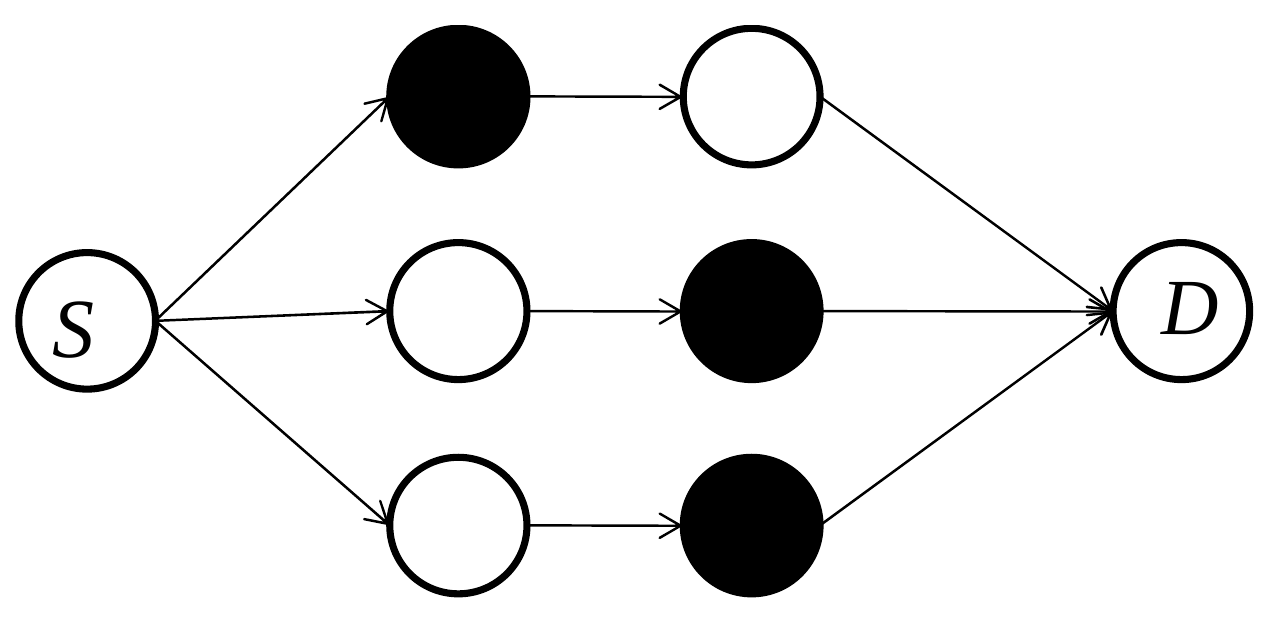}
\caption{An instance of a network with node-disjoint paths, with $n=3$ and $m=3$, the corresponding state is $S=(1,2,2)$.}
\label{fig:state}
\end{center}
\end{figure}

Consider a source S and its receiver D.
The network we study has $n$ paths and each path has $m$ hops.
The link error probability for each hop is equal to $e$.
The number of data packets that are coded together are $k$ (where $k \leq n$).
In order to find the average time that is needed for d to receive the packets, we model our problem using absorbing Markov Chains~\cite{b:prob}.
The chain is absorbed when the receiver d has received $k$ packets.
A state of this chain is denoted by $S$.
$S$ is a $n$- tuple: $S=(s_{1},s_{2},...,s_{n})$, where $s_{i}$ is the number of hops traversed by a packet on path $i$, note that $0 \leq s_{i} \leq m$ and $1 \leq i \leq n$.
For example in Figure~\ref{fig:state}, the nodes with black color are the ones that have received already the packet.

The state space denoted by $V_{S}$ contains all the $(m+1)^{n}$ states of the Markov Chain. $V_{S}$ is divided into two sub-spaces $V_{T}$ and $V_{A}$, $V_{S}=V_{T} \cup V_{A}$. $V_{T}$ and $V_{A}$ are the spaces that contain the transient and absorbing states respectively. There are $\left| V_{S} \right| = (m+1)^{n}$ states in total.
The absorbing ones are:
\begin{equation}
\left| V_{A} \right| = \sum_{i=k}^n {n\choose i}.
\end{equation}
The transient states are:
\begin{equation}
\left| V_{T} \right| =(m+1)^{n} - \sum_{i=k}^n {n\choose i}.
\end{equation}

The transition matrix $T$ of the Markov Chain has the following canonical form~\cite{b:prob}:

\begin{equation}
T=\left(\begin{array}{cc}P & R \\0 & I \end{array}\right).
\end{equation}

$P$ is an $\left| V_{T} \right|\times \left| V_{T} \right|$ matrix, $R$ is $\left| V_{T} \right|\times \left| V_{A} \right|$ and $I$ is $\left| V_{A} \right|\times \left| V_{A} \right|$ matrix. It is known that for an absorbing Markov Chain the matrix $I-P$ has an inverse~\cite{b:prob}. Also it is known that:
\begin{equation}
t=(I-P)^{-1}\textbf{1}_{\left| V_{T} \right|\times 1},
\end{equation}

where $t$ is the expected number of steps before the chain is absorbed and $\textbf{1}_{\left| V_{T} \right|\times 1}$ is the all-ones column vector. The first element of $t$ is the expected time for the chain to be absorbed starting from the initial state, that is the delay we want to compute.
The rest of this section presents the procedure in order to compute the matrix $P$.
We assign indices for the transient states, the initial state $S_{0}=(0,0,...,0)$ being the first one.
This indexing facilitates the computation of the elements of matrix $P$. For example $P_{ij}$ is the probability of transition from $S_{i}=(s_{1}^{i},...,s_{n}^{i})$ to $S_{j}=(s_{1}^{j},...,s_{n}^{j})$.
The elements of $P$ can be computed through the following equation:

\begin{equation}
\begin{aligned}
\scriptsize{
 P_{ij} = \left\{\begin{matrix}
0, & if \exists \text{ } k \text{ } s.t. \text{ } s_{k}^{j}<s_{k}^{i} \text{ } or \text{ } s_{k}^{j}-s_{k}^{i}>1\\
e^{n-cor-fin}(1-e)^{cor}, & otherwise.
\end{matrix}\right. }
\end{aligned}
\end{equation}

where

\begin{equation}
fin=\sum_{k=1}^n {\floor*{ \frac{s_{k}^i}{m} }},
\end{equation}

and

\begin{equation}
cor=\sum_{k=1}^{n}{(s_{k}^{j}-s_{k}^{i})}.
\end{equation}
The Markov Chain is absorbed when the receiver has received at least $k$ packets, which means $fin \geq k$.

Next we show how the previous procedure can be applied for the computation of the delay and throughput for single path, multipath, multicopy and multipath with network coding.

- Single Path: For this case, we apply the previous procedure with $n:=1$ and $k:=1$, to calculate the delay $D_{sp}$.
The throughput is given by $Thr_{sp}=\frac{1}{D_{sp}}$.

- Multipath: The delay for multipath is equal to $D_{sp}$.
The throughput is given by $Thr_{mp}=\frac{n}{D_{sp}}$.

- Multicopy: Multicopy is the technique for maximum redundancy, we send the same symbol to all paths.
We apply the previous procedure with $n:=n$ and $k:=1$, to calculate the delay for multicopy $D_{mcop}$.
The throughput is given by $Thr_{mcop}=\frac{1}{D_{mcop}}$.

- Multipath with Network Coding:
A packet generation of \textit{k} packets is assumed.
These packets are encoded together resulting in $n=2^{k}-1$ coded packets and one that contains only zero values.
Each coded packet is assigned on one of the $n$ paths.
The procedure is applied with parameter $n:=n$ and $k:=k$, to calculate the delay for network coding $D_{nc}$.
The throughput is given by $Thr_{nc}=\frac{k}{D_{nc}}$.

\subsubsection{Paths wtih Common Nodes, Hop-by-Hop Coding}
\label{sec:analysis_hopbyhop}

\begin{figure}[ht]
\centering
\subfigure[One hop]{
\includegraphics[scale=0.5]{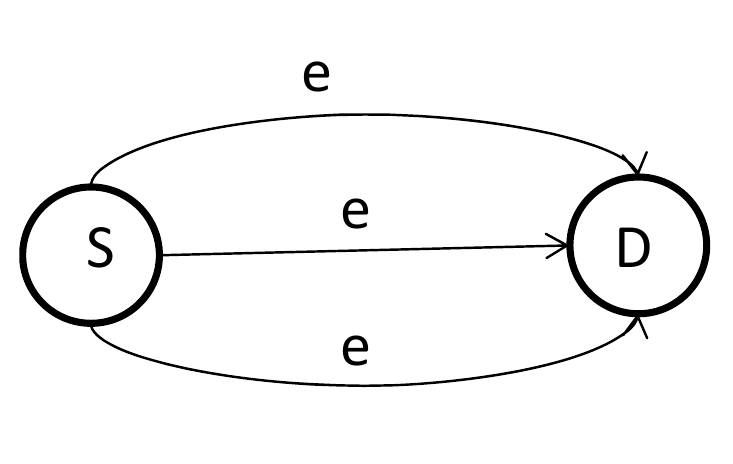}
\label{fig:onehop}
}
\subfigure[n hops]{
\includegraphics[scale=0.5]{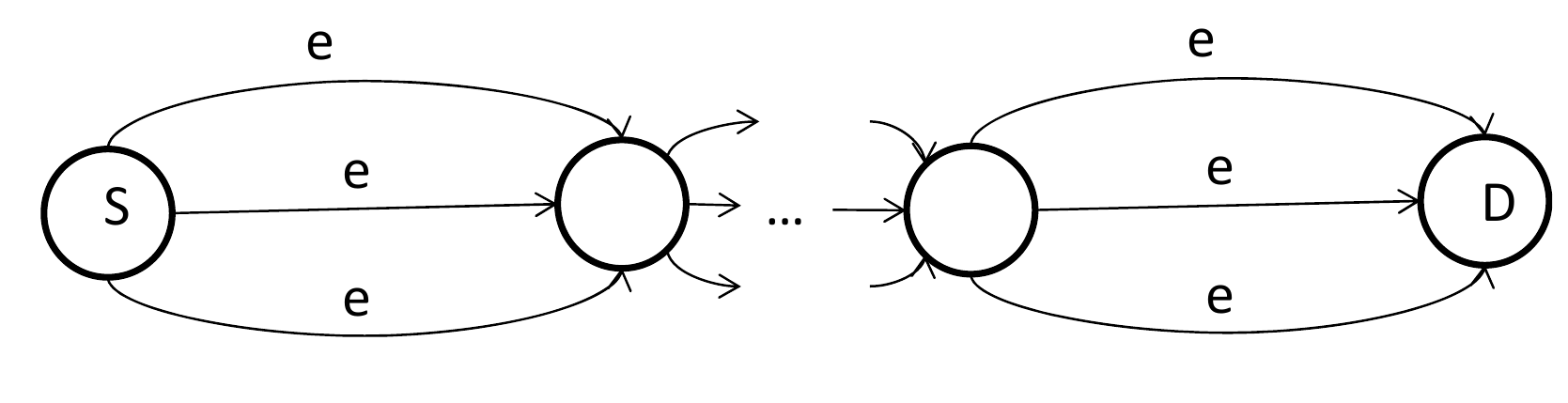}
\label{fig:n-hops}
}
\label{fig:network}
\caption{Simple network with three paths having nodes in common}
\end{figure}

The derivation of the equations in this section is based onsection II of \cite{b:pathdivgain1}.
There is a small change for the case of network coding.
We consider two different scenarios, one consisting of three paths and one consisting of seven both consisting of a single hop.
Moreover, the link error probability for each link-path is the same and equal to $e$.

\textit{A. Three Paths}

In this part we will present the equations corresponding to network depicted in Figure~\ref{fig:onehop}.

- Single Path: The average delay is given by $D_{sp}=\frac{1}{1-e} \,$ and the throughput is $Thr_{sp}=\frac{1}{D_{sp}}=1-e \, .$

- Multipath: Multipath has the same delay as the single path $D_{mp}=D_{sp} \,$ and its throughput is three times the throughput of single path $Thr_{mp}=3 Thr_{sp} \, .$

- Multicopy: The delay and throughput are $D_{mcop}=\frac{1}{1-e^{3}}$ and $Thr_{mcop}=\frac{1}{D_{mcop}}$ respectively.

- Multipath with Network Coding: The delay $D_{nc}$ is the average delay to receive at least two of the three independent linear combinations sent by node S:
$D_{nc}=\frac{(1-e)^{3}+3e(1-e)^{2}+3e^{2}(1-e)(1+D_{1})+e^{3}}{1-e^{3}}$
where $D_{1}=D_{mcop}=\frac{1}{1-e^{3}} .$
The additional delay $D_{1}$ is to receive one more linear combination when we have already received one.
Since in the time interval $D_{nc}$ node R receives two data packets, the average throughput is given by $Thr_{nc}=\frac{2}{D_{nc}} .$

\textit{B. Seven Paths}

- Single Path: The average delay is given by $ D_{sp}=\frac{1}{1-e}$ and the throughput is $Thr_{sp}=\frac{1}{D_{sp}}=1-e .$

- Multipath: Multipath has the same delay as the single path $D_{mp}=D_{sp} $ and its throughput is seven times the throughput of the single path $Thr_{mp}=7 Thr_{sp} .$

- Multicopy: The delay and throughput are $D_{mcop}=\frac{1}{1-e^{7}}$ and $Thr_{mcop}=\frac{1}{D_{mcop}} $ respectively.

- Multipath with Network Coding: We have three packets to transmit through $2^{3}-1=7$ paths.
According to lemma in appendix A in~\cite{b:pathdivgain1} we need at least three and at most four linear packet combinations to be able to decode the initial packets.
The delay for receiving three or four linear combinations is denoted by $D_{nc-L}$,$D_{nc-U}$ respectively.
\begin{align}
D_{nc-L}=\frac{1}{1-e^{7}}\left[\sum_{i=3}^{7} {{{7}\choose i}(1-e)^{i}e^{7-i}}
 + \sum_{i=1}^{2} {{{7}\choose i}(1-e)^{i}e^{7-i}(1+D_{3,3-i})}+e^{7} \right] \, ,
\nonumber
\end{align}
where $D_{3,i}$ is the delay to receive $i=1,2$ encoded packets when $3$ needed, $D_{3,1}=\frac{1}{1-e^{7}}$, $D_{3,2}=\frac{1}{1+e^{7}}[1-e^{7}+(1+\frac{1}{1-e^{7}})(e^{3}(1-e^{4})+e^{4}(1-e{^3})]$.
The average delay to receive $4$ linear combinations is given by:
\begin{align}
D_{nc-U}=\frac{1}{1-e^{7}}\left[\sum_{i=4}^{7} {{{7}\choose i}(1-e)^{i}e^{7-i}}+
 \sum_{i=1}^{3} {{{7}\choose i}(1-e)^{i}e^{7-i}(1+D_{4,4-i})}+e^{7}\right] \, ,
\nonumber
\end{align}
where $D_{4,i}$ is the delay to receive $i=1,2,3$ encoded packets when $4$ needed, $D_{4,1}=D_{3,1}$, $D_{4,2}=D_{3,2}$, $D_{4,3}=D_{nc-L}$.
The throughput is given by: $Thr_{nc}=\frac{3}{D_{nc}}$ .

\textbf{Remark:} If the network topology has $n$ hops as in figure~\ref{fig:n-hops}, then in order to find the total delay with the previous models we just need to add the delays for all the hops.
In the case where all links have the same error probabilities then the total delay is $n$ times the delay for one hop.

\subsubsection{Network with three single-hop paths with different link errors per hop}
\label{sec:diff_error_per_hop}

In this section we present the equations for the delay and throughput for the forwarding schemes discussed where link-paths have different error probabilities.
The derivation of the equations in this section is based on Appendix B of \cite{b:pathdivgain1}. There is a small change for the case of network coding.

- Single Path: This forwarding scheme selects the path with the lowest link-path error probability from the three available.
Thus the delay is $D_{sp}=\frac{1}{1-\min_{i}e_{i}}$ and the throughput is $Thr_{sp}=\frac{1}{D_{sp}}$.

- Multipath: In this routing scheme different data flows follow different paths, so the average delay per packet and the throughput are:
$D_{mp}=\frac{1}{3}\sum_{i=1}^{3}{\frac{1}{1-e_{i}}}$, $Thr_{mp}=\frac{3}{D_{mp}} $ respectively.

- Multicopy: The multicopy scheme employs all available paths to forward packets of a specific flow, achieving in this way the maximum redundancy (at the cost of wasted resources).
The average delay is: $D_{mcop}=1/(1-\prod_{i=1}^{3}{e_{i}})$ and the average throughput is: $Thr_{mc}=\frac{1}{D_{mc}}$.

- Multipath with Network Coding: Multipath with Network Coding uses all available paths sending linear combinations of initial packets (coded packets) to each of them.
In the topology examined three paths are available.
The packet generation assumed has a size of two so at least two coded packets are needed in order to decode the original data.
The average delay is given by:
\begin{align}
D_{nc}=\frac{1}{1-\prod_{i=1}^{3}{e_{i}}}\left[\prod_{i=1}^{3}{(1-e_{i})}+\sum_{i=1}^{3}{e_{i}\prod_{j=1, j \neq i}^{3}{(1-e_{j})}}
 + \sum_{i=1}^{3}{(1-e_{i})(1+D_{1})\prod_{j=1, j \neq i}^{3}{e_{j}}}+\prod_{i=1}^{3}{e_{i}}\right], \nonumber
\end{align}
where $D_{1}=\frac{1}{1-\prod_{i=1}^{3}{e_{i}}}$.
The throughput is: $Thr_{mc}=\frac{2}{D_{nc}}$.

\section{\uppercase{Numerical Results}}
\label{sec:numerical_results}

In this section we present numerical results drawn from the analytical framework and the network settings presented in the previous section.

\subsection{Node-disjoint Paths, End-to-End Coding, Same Link Error Probabilities}
\label{sec:numerical_node_disj}

\begin{table}[ht]
\centering
\begin{tabular}{c c c c c c}
\hline\hline
Scheme & Error & Paths & Hops & Delay/DelaySP & Thr/ThrSP\\ [0.5ex]
\hline
NC & 0.2 & 3 & 2 & 0.9312 & 2.148  \\
SP & 0.2 & 3 & 2 & 1 & 1  \\
MP & 0.2 & 3 & 2 & 1 & 3 \\
MC & 0.2 & 3 & 2 & 0.819 & 1.221 \\ [1ex]

NC & 0.2 & 3 & 4 & 0.967 & 2.07  \\
SP & 0.2 & 3 & 4 & 1 & 1  \\
MP & 0.2 & 3 & 4 & 1 & 3 \\
MC & 0.2 & 3 & 4 & 0.845 & 1.184 \\ [1ex]

NC & 0.4 & 3 & 2 & 0.93 & 2.15  \\
SP & 0.4 & 3 & 2 & 1 & 1  \\
MP & 0.4 & 3 & 2 & 1 & 3 \\
MC & 0.4 & 3 & 2 & 0.694 & 1.44 \\ [1ex]

NC & 0.4 & 3 & 4 & 0.967 & 2.07  \\
SP & 0.4 & 3 & 4 & 1 & 1  \\
MP & 0.4 & 3 & 4 & 1 & 3 \\
MC & 0.4 & 3 & 4 & 0.761 & 1.31 \\ [1ex]

NC-L & 0.2 & 7 & 2 & 0.825 & 3.64  \\
NC-U & 0.2 & 7 & 2 & 0.888 & 3.38  \\
SP & 0.2 & 7 & 2 & 1 &  1  \\
MP & 0.2 & 7 & 2 & 1 & 7 \\
MC & 0.2 & 7 & 2 & 0.8 & 1.25 \\ [1ex]

NC-L & 0.4 & 7 & 2 & 0.771 & 3.89  \\
NC-U & 0.4 & 7 & 2 & 0.903 & 3.32  \\
SP & 0.4 & 7 & 2 & 1 &  1  \\
MP & 0.4 & 7 & 2 & 1 & 7 \\
MC & 0.4 & 7 & 2 & 0.613 & 1.63 \\ [1ex]
\hline
\end{tabular}
\caption{Delay and Throughput Tradeoff for all forwarding schemes}
\label{table:trnodedisjoint}
\end{table}

Table \ref{table:trnodedisjoint} presents the throughput and delay for all forwarding schemes for the case of node disjoint paths where links share the same error probability.
As far as network coding is concerned, end-to-end  coding is assumed.
Three different topologies are considered based on the number of paths and number of hops per path.
As this table shows, for the scenario with three paths with two hops each, multipath with network coding achieves delay which is smaller than single and multipath, but worst than multi-copy forwarding.
The gain is approximately $7\%$ for link error probabilities $0.2$ and $0.4$.
As far as throughput is concerned, the throughput achieved by multipath with network coding is better than that achieved by multicopy forwarding.
It is also interesting to note that when each path consists of four hops instead of two the gain of network coding in terms of delay decreases approaching multipath's delay.
This is expected because of the relatively small number of paths and packets.

Concerning the topology consisting of seven paths and two hops, table \ref{table:trnodedisjoint} includes two entries for network coding, one corresponding to the
case where the receiver is able to decode a packet generation after receiving three linear combinations (which is denoted by NC-L) and one for decoding after having received four (which is denoted by NC-U);
These number represent the lower and upper bound of the number of coded packets required to retrieve all packets at the receiver, as indicated by lemma in \cite{b:pathdivgain1}.
Multipath with network coding (NC-U) achieves delay, which is better than single and multipath (about $11\%$ for $e=0.2$ and $9.7\%$ for $e=0.4$), but worst than multicopy forwarding.
In terms of throughput, network coding (NC-U) performs much better than multicopy achieving $170\%$ and $103.6\%$ higher throughput for $e=0.2$ and $e=0.4$ respectively.
Multicopy is superior for high error probabilities and for a large number of hops because of its higher redundancy.
Throughput achieved by multipath with network coding is better than that achieved by multi-copy routing.
Further on, multipath with network coding outperforms multicopy in terms of throughput.

\subsection{Paths with Common Nodes, Hop-by-Hop Coding, Same Link Error Probabilities}
\label{sec:numerical_hopbyhop}

\begin{table}[ht]
\centering
\begin{tabular}{c c c c c c}
\hline\hline
Scheme & Error & Paths & Delay/DelaySP & Thr/ThrSP\\ [0.5ex]
\hline
NC & 0.2 & 3 & 0.8845 & 2.261  \\
SP & 0.2 & 3 & 1 & 1  \\
MP & 0.2 & 3 & 1 & 3 \\
MC & 0.2 & 3 & 0.807 & 1.24 \\ [1ex]

NC & 0.4 & 3 & 0.838 & 2.386  \\
SP & 0.4 & 3 & 1 & 1  \\
MP & 0.4 & 3 & 1 & 3 \\
MC & 0.4 & 3 & 0.641 & 1.56 \\ [1ex]

NC-L & 0.2 & 7 & 0.804 & 3.733  \\
NC-U & 0.2 & 7 & 0.827 & 3.629  \\
SP & 0.2 & 7 & 1 &  1  \\
MP & 0.2 & 7 & 1 & 7 \\
MC & 0.2 & 7 & 0.8 & 1.25 \\ [1ex]

NC-L & 0.4 & 7 & 0.656 & 4.573  \\
NC-U & 0.4 & 7 & 0.777 & 3.862  \\
SP & 0.4 & 7 & 1 &  1  \\
MP & 0.4 & 7 & 1 & 7 \\
MC & 0.4 & 7 & 0.601 & 1.664 \\ [1ex]
\hline
\end{tabular}
\caption{Delay-Throughput Tradeoff for paths with node in common}
\label{table:trcommon}
\end{table}

Table~\ref{table:trcommon} shows the throughput-delay tradeoff for a network with paths having nodes in common for error probabilities $e=0.2$ and $e=0.4$.
For the case of three paths, multipath with network coding achieves delay, which is better than single and multipath (approximately $11.5\%$ and $16\%$ for e = 0.2 and e = 0.4 respectively),
but worst than multi-copy forwarding.
In terms of throughput, network coding is much better($82.3\%$ and $52.9\%$ for e = 0.2 and e = 0.4 respectively) than multicopy.
For the case of seven paths, network coding (NC-U) achieves delay, which is better than single and multipath (about $17\%$ and $22\%$ for e = 0.2 and e = 0.4 respectively), but slightly worse than multicopy forwarding.
In terms of throughput network coding (NC-U) is much better than multicopy ($190.4\%$ and $132\%$ for $e=0.2$ and $e=0.4$ respectively).

\subsection{Network with three single-hop paths with different link errors per hop}
\label{sec:numerical_diff_error_per_hop}

Table~\ref{table:trdiff} shows the delay-throughput trade-off for two different scenarios.
\begin{table}[ht]
\centering
\begin{tabular}{c c c c c c}
\hline\hline
Scheme & $e_{1}$ & $e_{3}$ & $e_{2}$ & Delay/DelaySP & Thr/ThrSP\\ [0.5ex]
\hline
NC & 0.3 & 0.4 & 0.5 & 0.974 & 2.053  \\
SP & 0.3 & 0.4 & 0.5 & 1 & 1  \\
MP & 0.3 & 0.4 & 0.5 & 1.189 & 2.523 \\
MC & 0.3 & 0.4 & 0.5 & 0.745 & 1.343 \\ [1ex]

NC & 0.5 & 0.6 & 0.8 & 1.056 & 1.894  \\
SP & 0.5 & 0.6 & 0.8 & 1 & 1  \\
MP & 0.5 & 0.6 & 0.8 & 1.583 & 1.895 \\
MC & 0.5 & 0.6 & 0.8 & 0.658 & 1.52 \\ [1ex]

\hline
\end{tabular}
\caption{Delay-Throughput Tradeoff for three paths with different error probabilities}
\label{table:trdiff}
\end{table}

In the case of $e_{1}=0.5$, $e_{3}=0.6$ and $e_{2}=0.8$ network coding is the superior forwarding scheme and has almost the same delay as the single path but the double throughput.
Multipath has the same throughput with network coding but $58\%$ higher delay than single path.

Summarizing the above we can state that network coding offers significant advantages as the number of paths increases, when the nodes inside the network are able to decode and encode the received packets.

\section{\uppercase{Simulation Setup and Results}}
\label{sec:simulation}

In section \ref{sec:sim_setup} we describe the simulation setup and parameters along with the scenarios simulated while in section \ref{sec:sim_result} the simulation results are presented.

\subsection{\uppercase{Simulation Setup}}
\label{sec:sim_setup}
We evaluate the throughput and delay of all aforementioned forwarding schemes  using network simulator NS-2, version 2.34 \cite{ref:ns2}, including support for multiple transmission rates and an SINR-based packet
level error model \cite{ref:dei80211mr}.
A custom source-routed routing protocol is employed to ensure that packets of the same flow are forwarded through the same path.
Traffic sources employ static predefined routes to the destination and generate constant bit rate UDP flows.
Implementing a search algorithm for node-disjoint paths is out of the scope of the evaluation process.
Concerning medium access control, a slotted aloha-based MAC layer is implemented.
Transmission of data, routing protocol control and ARP packets is performed at the beginning of each slot without performing carrier sensing prior to transmitting.
Acknowledgments for data packets are sent immediately after successful packet reception while failed packets are retransmitted.
Slot length $T_{slot}$ is expressed through: $T_{slot} = T_{data} + T_{ack} + 2D_{prop}$ where $T_{data}$ and $T_{ack}$ denote the transmission times for data packets and ACKs while $D_{prop}$ denotes the propagation delay.
It should be noted that all packets have the same size shown in table \ref{tab:param_simul}.
All network nodes, apart from sources of traffic, select a random number of slots before transmitting drawn uniformly from $[0,CW]$ where contention window (CW) is fixed for the whole duration of the simulation and equal to 7.
On the long term and adopting the previous assumptions concerning the contention window, if there is always a packet available for transmission at the queues of the relay nodes then for each relay node approximately 22.2\% of the slots will be occupied for packet transmission
with the rest of them spent either for receiving or being idle.
For sources of traffic, the transmission probability is fixed in order to control the rate at which traffic is injected into the network with sources of traffic denoting different
interfaces of a single node.
We explore three different scenarios concerning the transmission probability of traffic sources:
\begin{itemize}
\item Lower than the transmission probability of relay nodes and equal to 0.1
\item Almost equal with the transmission probability of relay nodes and equal to 0.2
\item Higher and equal to 0.3.
\end{itemize}
Due to space limitations results for transmission probability equal to $0.2$ are presented in the rest of the section.
Simulation results for other transmission values are presented in \ref{sec:appendix}.
Additionally, all nodes share the same channel, transmission rate, and power (parameter values summarized in table ~\ref{tab:param_simul}).
As far as queue size at each node is concerned it is set to a sufficiently large value so that no packet is dropped due to buffer overflow during the simulation period.

\begin{table}[t]
\caption{Parameters used in the simulations}
\label{tab:param_simul}
\begin{center}
\begin{tabular}{ll}
\hline
Parameter & Value\\ \hline
RTS/CTS & Off\\
Max Retransmit Threshold & Off\\
Link Rate & 24Mbps\\
Transmit Power (EIRP) & 20 dBm\\
Propagation Model & Freespace\\
System Loss & 0 dBm\\
Contention Window & 7\\
Packet payload + UDP Header & 1500 Bytes\\
\hline
\end{tabular}
\end{center}
\end{table}

As also described in our previous work \cite{6335387}, adding support for simulating network coding requires two main modifications. Firstly, data packets that are coded together
and thus belong to the same generation are marked with a common \textit{generation id}. In this way, receivers are able
to distinguish among packets from different generations and decode them.
The second modification concerns the assumption introduced in our prior work \cite{ref:papas_hop_by_hop} according to which relay nodes remove from their queues a multi-copied packet that is successfully delivered to the destination or any
packets that belong to a generation that is successfully decoded by the destination. To support this functionality,
a \textit{global ack} mechanism is simulated which consists of a custom acknowledgment broadcasted throughout the whole network by the destination node upon
reception of a packet or successful decoding
of a packet generation.
This acknowledgment carries the sequence number of the packet received for the case of multicopy and the generation id of
the generation decoded for the case of network coding-based forwarding.

In each simulated scenario the source node generates a flow $f$ of $R=9Mbps$ constant bit rate UDP traffic consisting of 1500 bytes packets routed to the destination over $n$ multiple paths in parallel.
Mulipath splits $f$ into $n$ subflows of rate $R_{i}=R/n, i=1...n$.
Each subflow is forwarded to the destination through a specific interface of the source node and a predefined path.
Multicopy replicates $f$ on all paths assigning a subflow of rate $R_{i}=R$ on each one.
For the case of network coding, assuming a packet generation of size $k$ (number of data packets coded together), a subflow or rate $R_{i}=R/k$ is assigned on each path.
Single path on the other hand routes $f$ to the destination through the shortest path available to it.

It should also be noted that in the simulated scenarios we explore two basic variants of network coding.
Following the assumptions of the analytical framework presented in section \ref{sec:analysis} the first variant
of network coding explored allows only one packet generation to be on the network each time.
Subsequent packet generations are injected into the network only when the previous one is fully decoded at the destination.
For the rest of the study the notation used for this variant will be \textit{NC} or \textit{NC-L} and \textit{NC-U} for the case of seven paths (also explained in section \ref{sec:numerical_results}).
The second network coding variant explored is a greedy one that continually injects packet generations into the network without waiting for the previous ones to be decoded.
For the rest of the study this variant will be referred to as \textit{G-NC} or \textit{G-NC-L} and \textit{G-NC-U} for scenarios consisting of seven paths.

\begin{figure}
\begin{center}
\includegraphics[scale=0.4]{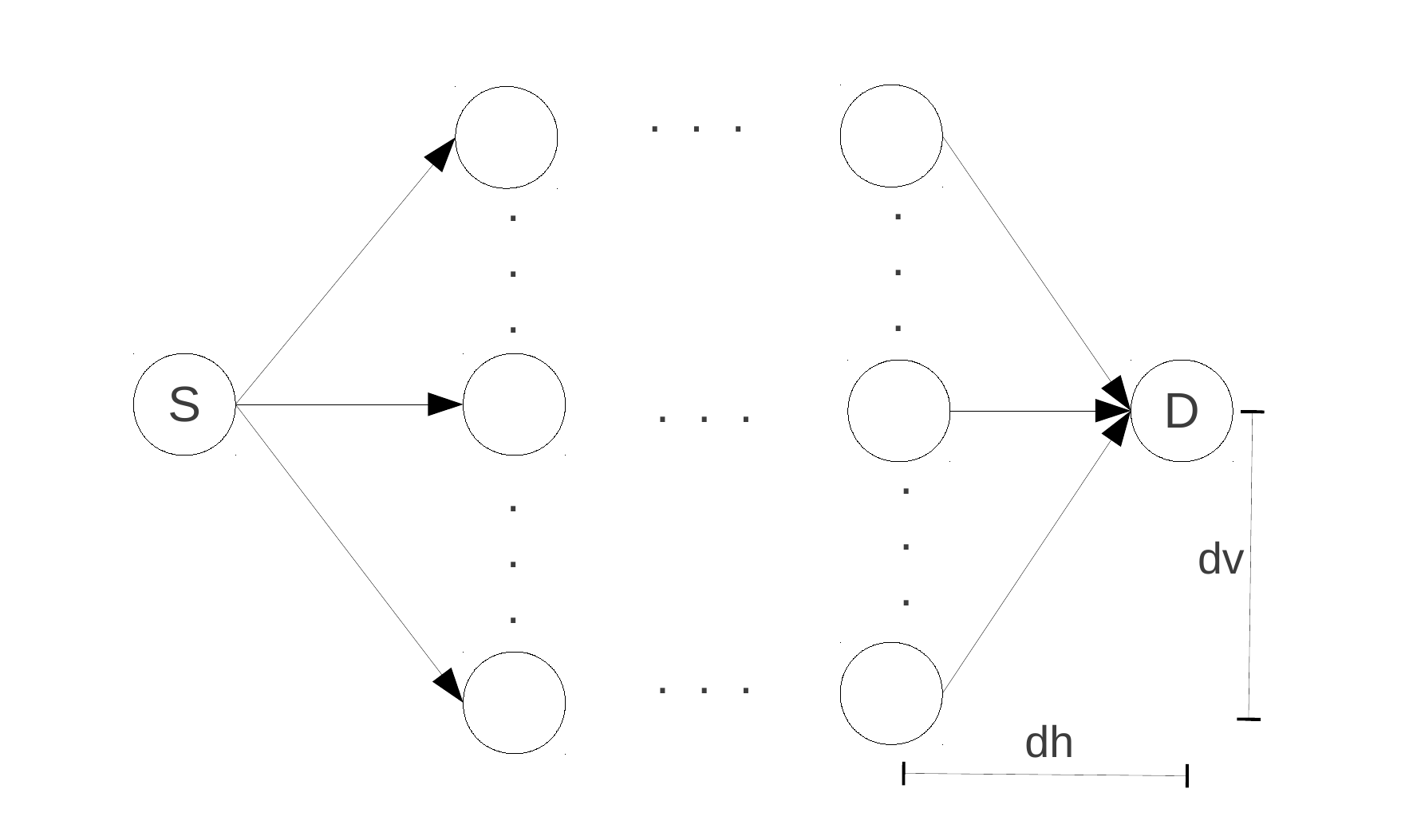}
\caption{Wireless topology of $n$ paths with $m$ hops each with equal vertical and horizontal distance between relays}
\label{fig:generic_topo_1}
\end{center}
\end{figure}

All the simulation results presented in the rest of the section are drawn from two different types of topologies.
The first type presented in figure \ref{fig:generic_topo_1} consists of multiple multi-hop paths.
The vertical distance between any two neighboring relay nodes is $d_{v}$ meters while the horizontal distance between any two nodes in the same path is $d_{h}$ meters.
An example of such a topology for three paths with three hops each is depicted in figure \ref{fig:state}.
The second type of topology consists of multiple single-hop paths where the source and destination node are assumed to be $d_{h}$ meters apart.
An example of such a topology with three single hop paths is depicted in figure \ref{fig:onehop}.
For each different topology employed all the aforementioned forwarding schemes are employed resulting in one simulation scenario for each pair of
topology and forwarding scheme.
For reasons of fair comparison among the schemes evaluated, each simulated scenario is considered finished when the receiver successfully receives or decodes $2000$ packets.
Tables \ref{tab:sim_topol_1} to \ref{tab:sim_topol_5} below present simulation results for the topologies for which numerical results were extracted in section \ref{sec:numerical_results}.

\subsection{\uppercase{Simulation Results}}
\label{sec:sim_result}

Before presenting simulation results a brief discussion about how delay and throughput are measured for each scheme is provided.
For the case of single path (SP) and multipath (MP), delay is estimated as the average per-packet delay with per-packet delay denoting the time
interval between dequeuing a packet for transmission at source node S and successful reception of that packet at destination D.
As far as multicopy (MC) is concerned, delay is also estimated as average per-packet delay. However, in this case, per-packet delay denotes the interval
between dequeuing a packet with sequence number $k$ at S and the time when the first packet with sequence number $k$ is received at D.
In case of network coding based schemes and assuming a packet generation of size $n$, delay is estimated as average per-generation delay where per-generation delay is the interval between dequeuing
the first coded packet of a specific packet generation $i$ at source node S and the time when destination D receives the $n^{th}$ coded packet for that generation.
Recall that destination D is able to decode a generation when it receives at least $n$ coded packets of that generation.
For the case of network coding based schemes, \textit{inter-arrival times} reports the average inter-arrival time over all coded packets of all generations received at the destination
with inter-arrival time denoting the interval between the successful reception of two successive coded packets at the receiver.
Finally the row labeled \textit{Pkt drops} presents the total number of data (or coded for the case of network coding) packets that are dropped due to noise and interference.

\begin{table}[t]
\begin{center}
\scriptsize
\begin{tabular}{|l|c|c|c|c|c|}
\hline
~                                    & MP & MC & NC & G-NC & SP \\ \hline
Delay (msec)   & 198.7   & 75.8  & 26.7  & 329.4 & 49.2  \\ \hline
Throughput                   & ~  & ~  & ~  & ~         & ~  \\
(Mbps)                   & 1.81  & 1.02  & 1.70 & 1.50 & 2.14 \\ \hline
Inter-arrival              & ~  & ~  & ~  & ~         & ~  \\
times (msec)   &   &   &  7.8 & 127.3 & \\ \hline
Pkt drops  & 28342  & 36981  & 12328  & 30293 & 3386 \\ \hline
\end{tabular}
\end{center}
\caption {Three paths, four hops. $d_{h}=40m$, $d_{v}=80m$}
\label{tab:sim_topol_1}
\end{table}

\begin{table}[t]
\begin{center}
\scriptsize
\begin{tabular}{|l|c|c|c|c|c|}
\hline
~                                    & MP & MC & NC & G-NC & SP \\ \hline
Delay (msec)   & 205.6   & 94.7  & 29.8  & 6215.0 & 49.2  \\ \hline
Throughput                   & ~  & ~  & ~  & ~         & ~  \\
(Mbps)                   & 1.98  & 1.41  & 1.52 & 1.07 & 2.14 \\ \hline
Inter-arrival              & ~  & ~  & ~  & ~         & ~  \\
times (msec)   &   &   &  12.2 & 2727.3 & \\ \hline
Pkt drops  & 22701  & 19857  & 13307  & 37652 & 3386 \\ \hline
\end{tabular}
\end{center}
\caption {Three paths, four hops. $d_{h}=40m$, $d_{v}=120m$}
\label{tab:sim_topol_2}
\end{table}

\begin{table}
\begin{center}
\scriptsize
\begin{tabular}{|l|c|c|c|c|c|c|c|}
\hline
~                                   & MP & MC & NC-L & NC-U & G-NC-L & G-NC-U & SP \\ \hline
Delay           &   &   &  &  & & & \\
(msec)           & 397.0  & 39.4  & 46.9 &  49.7   &  380.1 &  446.9 & 21.6\\ \hline
Throughput             & ~  & ~  & ~    & ~    & ~           & ~           & ~  \\
(Mbps)                 & 1.52  & 0.65  &  1.46   &  1.39  & 1.12  & 0.90 & 3.06  \\ \hline
Inter-arrival             & ~  & ~  & ~    & ~    & ~           & ~           & ~  \\
times (msec)   &   &   &  19.2 & 26.7 &  112.2 &  248.3 &   \\ \hline
Pkt drops          & 61218  & 103649  & 41785 & 41154 & 69925 & 92947 & 478 \\ \hline

\end{tabular}
\end{center}
\caption {Seven paths, two hops. $d_{h}=40m$, $d_{v}=10m$}
\label{tab:sim_topol_3}
\end{table}

\begin{table}[t]
\begin{center}
\scriptsize
\begin{tabular}{|l|c|c|c|c|c|}
\hline
~                                    & MP & MC & NC & G-NC & SP \\ \hline
Delay           &   &   &  &  & \\
(msec)   & 5.0   & 2.3  & 7.0  & 13.6 & 3.1  \\ \hline
Throughput                   & ~  & ~  & ~  & ~         & ~  \\
(Mbps)                   & 7.06  & 3.8  & 5.12 & 5.67 & 3.69\\ \hline
Inter-arrival              & ~  & ~  & ~  & ~         & ~  \\
times (msec)   &   &   &  4.18 & 9.7 & \\ \hline
Pkt drops  & 1299  & 2224  & 1389  & 1492 & 30 \\ \hline
\end{tabular}
\end{center}
\caption {Three paths, one hop. $d_{h}=40m$}
\label{tab:sim_topol_4}
\end{table}

\begin{table}
\begin{center}
\scriptsize
\begin{tabular}{|l|c|c|c|c|c|c|c|}
\hline
~                                   & MP & MC & NC-L & NC-U & G-NC-L & G-NC-U & SP \\ \hline
Delay           &  11.5 &  3.39 & 14.1 & 16.2 & 27.0 & 47.1 & 3.1\\
(msec)           &   &  &   &   &   &  &  \\ \hline
Throughput             & ~  & ~  & ~    & ~    & ~           & ~           & ~  \\
(Mbps)                 & 7.12 & 2.32  & 4.20 & 3.85 & 4.28 & 3.77 & 3.69\\ \hline
Inter-arrival             & ~  & ~  & ~    & ~    & ~           & ~           & ~  \\
times (msec)   &   &   & 8.7 & 11.6 & 19.0 & 37.7 &  \\ \hline
Pkt drops       & 6952 & 17263 &  8600 & 8595  & 9225 & 10461 & 30 \\ \hline
\end{tabular}
\end{center}
\caption {Seven paths, one hop. $d_{h}=40m$}
\label{tab:sim_topol_5}
\end{table}

\begin{figure}[ht]
\centering
\subfigure[Delay vs number of hops]{
\includegraphics[scale=0.4]{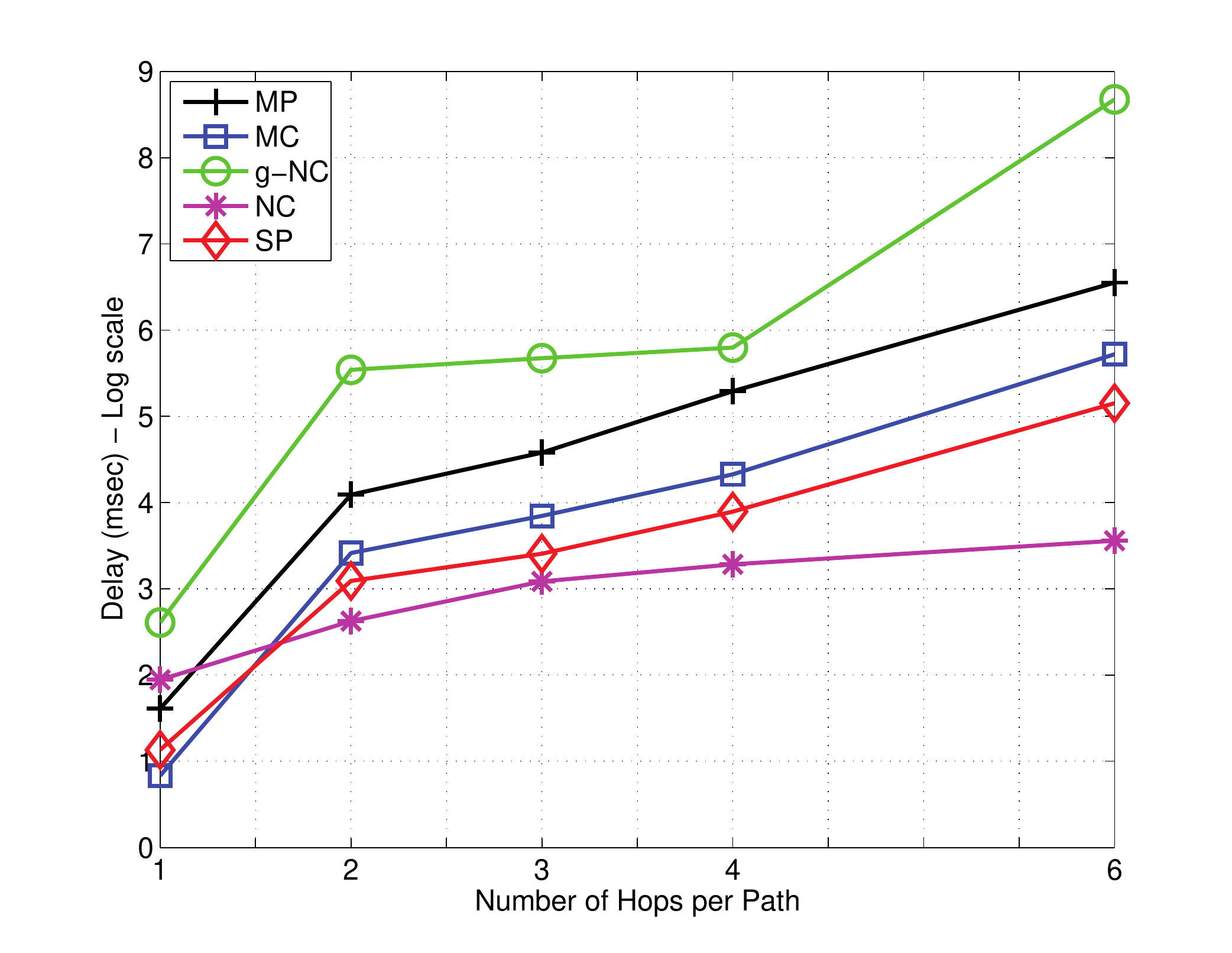}
\label{fig:num_hops_effct_delay}
}
\subfigure[Throughput vs number of hops]{
\includegraphics[scale=0.4]{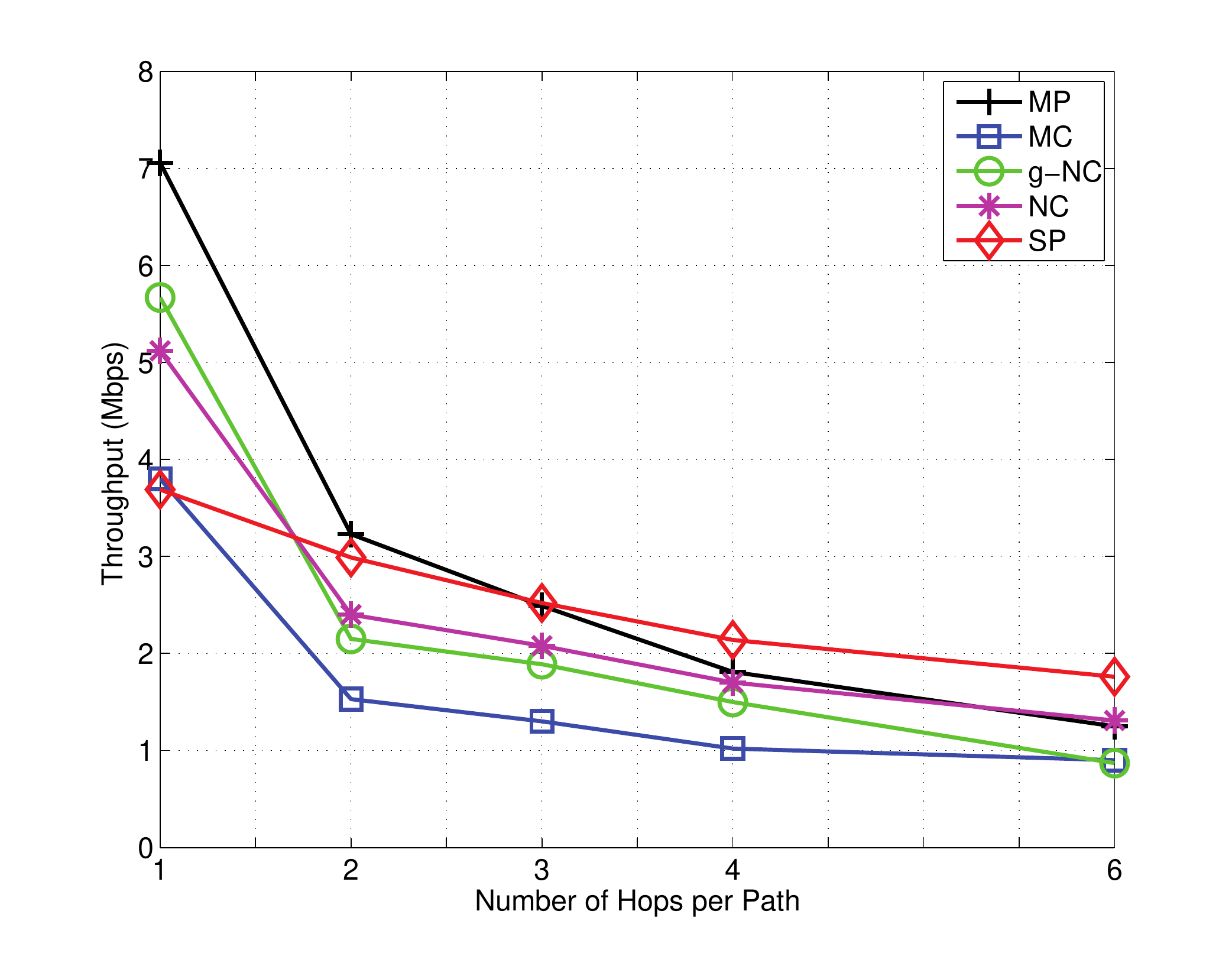}
\label{fig:num_hops_effct_thrput}
}
\label{fig:num_of_hops_eff}
\caption{Delay and throughput for a different number of hops, in the case of three paths and $e=0.2$ (node disjoint paths)}
\end{figure}

In order to explore the effect of the number of hops per path and the throughput and delay achieved we employ three more topologies consisting all of three paths and two, three, and six hops respectively.
Each of the aforementioned forwarding schemes is employed for each topology resulting in a new simulation scenario that is also considered finished when the receiver successfully receives or decodes $2000$ packets.
Figures \ref{fig:num_hops_effct_delay} and \ref{fig:num_hops_effct_thrput} present the delay and throughput achieved by each scheme on each topology.

As figure \ref{fig:num_hops_effct_delay} shows, the scheme that experiences the lower increase in terms of delay when the number of hops per path increases is network
coding (NC) with SP coming next.
An important reason for this is the fact that longer paths also imply higher intra- and inter-path interference.
As far as throughput is concerned, SP experiences the lower decay when the number of hops increases.

\section{\uppercase{Discussion}}
\label{sec:discussion}

In this section we compare the throughput and delay trends observed in the simulation results and the numerical ones drawn from our analytical framework for the topologies already described.
Our main goal is to validate and extend the trends in terms of delay and throughput revealed by our analytical framework.
It should be noted that it is not our intention to directly compare the throughput and delay values for the numerical and simulation results since the simulated scenarios are based on realistic wireless settings.
On the other hand our analysis disregards queuing delay and the waiting times introduced by back-off procedure at each transmitter.
Tables \ref{tab:comp_1} to \ref{tab:comp_4} contradict simulation and theoretical results for the four topologies explored in section \ref{sec:numerical_results}.
As far as \textit{rank} is concerned in these tables, the lower the rank of a scheme, the lower its delay and the higher its throughput.

\begin{table}[h]
\begin{center}
\scriptsize
\begin{tabular}{|c|c|c|c|c|c|c|c|}
\hline
     & \multicolumn{4}{|c|}{Simulation}    & \multicolumn{2}{|c|}{Numerical}  \\ \hline
     & \multicolumn{2}{|c|}{$d_{h}=40m$, $d_{v}=80m$} & \multicolumn{2}{|c|}{$d_{h}=40m$, $d_{v}=120m$}  & \multicolumn{2}{|c|}{Error=\{$0.2, 0.4$\}}  \\ \hline

Rank & Delay                  & Throughput           & \begin{tabular}[c]{@{}c@{}} Delay\\
\end{tabular}
& Throughput & Delay            & Throughput \\ \hline
1 & NC  & SP  & NC   & SP  & MC &  MP \\ \hline
2 & SP  & MP  & SP   & MP  &  NC & NC  \\ \hline
3 & MC  & NC  & MC   & NC  &  SP,MP & MC  \\ \hline
4 & MP  & G-NC  & MP   & MC  &  & SP \\ \hline
5 & G-NC  & MC  & G-NC & G-NC &  &   \\ \hline
\end{tabular}
\end{center}
\caption {Numerical vs simulation results. Three paths  four-hops paths}
\label{tab:comp_1}
\end{table}

Table \ref{tab:comp_1} contradicts the throughput and delay trends for the numerical results presented in table \ref{table:trnodedisjoint} and the simulation results of tables \ref{tab:sim_topol_1},\ref{tab:sim_topol_2}
for the case of a network consisting of three node-disjoint paths with four hops each.
Moreover, end-to-end coding is assumed for network coding.

We first consider the simulation topology where the vertical distance ($d_{v}$) between relay nodes is $80$ meters suggesting the presence of significant inter-path interference.
As far as delay is concerned, network coding (NC) achieves lower delay than single path (SP) and multipath (MP) in both the simulation and the numerical results.
Moreover, multicopy (MC) also achieves lower delay than MP.
What differs in the simulation results is the rank of MC in terms of delay which is higher than the delay achieved of both SP and NC.
The main reason for that is the effect of inter- and intra-path interference which is more prominent in the case of MC that continually utilizes all paths by forwarding high rate flows over them.
Recall that NC defers injecting the next packet generation into the network until the previous one is successfully decoded avoiding thus interference that would be caused by
transmission of coded packets belonging to other packet generations.
SP on the other hand utilizes only one path to the destination and thus suffers from only intra-path interference.
It is also interesting to note that the delay achieved by NC is lower than SP.
This is due to queuing delay which is more prominent in the case of SP that forwards the whole traffic of a high rate flow through a single path as opposed to NC that splits the traffic among the available paths.
According to the simulation setup presented in section \ref{sec:sim_setup}, SP forwards a flow of $9$ Mbps while NC splits the main flow into three subflows of $3$ Mbps each for a topology consisting of three paths.
To validate this observation, the \textit{queue occupation ratio} was estimated for all nodes in the two different simulation scenarios where SP and NC-based forwarding were employed
with queue occupation ratio being defined as the number of packets stored in the queue over the size of the queue in terms of number of packets.
For the case of SP forwarding, the queue occupation ratio was 0.72 for the first relay node while for NC the corresponding value for the first relay node of each path was 0.39.
This shows that packets forwarded through SP scheme wait for more slots in the queue before being transmitted.
The greedy variant of network coding (G-NC) suffers the highest delay of all schemes.
Successive injection of packet generations into the network without waiting for the previous ones to be decoded results in high interference and thus to increased number of retransmission required in order
to accomplish the transmission of a coded packet.
For the case of G-NC this delay is aggregate over all the coded packets that the destination must wait for in order to decoded a packet generation.

As far as throughput is concerned, our analytical framework underestimates the rank on only single path (SP).
The simulation results reveal that schemes that avoid inter-path interference are favored in terms of throughput. SP and NC thus outperform MP, MC and the greedy variant of network coding
with SP achieving the best throughput-delay trade-off.

When the vertical distance becomes $120$ meters, the effect of interference is alleviated.
However, the quality of the first and last links of the two outer paths in the topology depicted in figure \ref{fig:generic_topo_1} deteriorates due to higher distance.
Indeed, the number of packets dropped in this simulated scenario due to low SNR are $6428$ while the corresponding value for the scenario where the vertical distance between relays is $80$ meters is $549$ packets.
Still however, the receiver of the first link of the path in the middle faces significant interference from packets transmitted on the first links on the two outer paths.
In this simulated topology, lowest delay is achieved by SP and NC.
The reason is that these schemes avoid utilizing the lowest quality paths including the long distance links.
SP utilizes the shortest path of the three to the destination while NC distributes uniformly traffic among paths with high and lower quality also reducing inter-path interference.
Although MC injects high rate flows on all paths and thus causes significant interference, it achieves the next lowest delay since it delivers $93.6$\% of the packets to the destination
through the shortest path (the middle one).
MP on the other hand does not moderate the utilization of the two outer paths resulting in some packets experiencing large delay.
It is also interesting to note that when the vertical distance between relays becomes larger and thus the interference experienced decreases, MP manages to achieve higher
throughput than NC due to its lower redundancy.

\begin{table}[h]
\begin{center}
\scriptsize
\begin{tabular}{|c|c|c|c|c|}
\hline
     & \multicolumn{2}{|c|}{Simulation}     & \multicolumn{2}{|c|}{Numerical}                                   \\ \hline
     & \multicolumn{2}{|c|}{$d_{h}=40m$, $d_{v}=10m$} & \multicolumn{2}{|c|}{Error=\{0.2, 0.4\}} \\ \hline
Rank & Delay               & Throughput              & Delay            & Throughput            \\ \hline
1    & SP                 & SP                & MC              & MP          \\ \hline
2    & MC                 & MP                & NC-L              & NC-L      \\ \hline
3    & NC-L                & NC-L                & NC-U              & NC-U   \\ \hline
4    & NC-U                 & NC-U                & SP,MP              & MC   \\ \hline
5    & G-NC-L                 & G-NC-L                &               & SP     \\ \hline
6    & MP                 & G-NC-U                &               &           \\ \hline
7    & G-NC-U                 & MC                &               &           \\ \hline
\end{tabular}
\end{center}
\caption {Numerical vs simulation results. Seven two-hop paths}
\label{tab:comp_2}
\end{table}

Table \ref{tab:comp_2} contradicts the throughput and delay trends for the numerical results presented in table \ref{table:trnodedisjoint} and the simulation results of table \ref{tab:sim_topol_3},
for the case of a network consisting of seven node-disjoint paths with two hops each.
As far as network coding based schemes are concerned end-to-end coding is assumed.

Comparing the numerical with the simulation results, we observe that our analytical framework captures the rank in terms of delay for the various forwarding schemes apart from the case of SP.
SP achieves lower delay than all schemes employing multiple paths, although it experiences large queuing delay (discussed for the previous topology) since it avoids inter-path interference.
Indeed, as table \ref{tab:sim_topol_3} shows, SP experiences a dramatically lower number of dropped packets due to path loss and interference than all other schemes.
This is also why SP achieves higher throughput than all other forwarding schemes.
It is also interesting to note that although the high interference imposed on the network, MC achieves lower delay than NC.
More interestingly, MC suffers $148$\% more packet losses than NC due to path loss or interference.
One reason for NC's higher delay is the inter-arrival times aggregated over all coded packets that will be used by the destination to decode a specific packet generation.
For the scenario discussed the mean inter-arrival time for any pair of coded packets is $19.2$msec while at least three coded packets are needed to decode a packet generation.
The average per packet delay for MC is $39.4$msec.
The second reason for multicopy achieving lower delay than NC is the higher redundancy employed.
It should also be noted that MC achieves lower delay than schemes that do not mitigate interference, such as, MP and greedy network coding based schemes (G-NC-L, G-NC-U).
It is also interesting to note that although MC experiences a larger number of dropped frames due to path loss and interference than all other schemes employing multiple paths,
their effect on delay is balanced by the gain due to higher redundancy.
This is also obvious for the case of G-NC-L and MP where G-NC-L achieves lower delay than MP although its higher number of dropped frame due to interference.

As far as throughput is concerned, in accordance with the numerical results, MP achieves the highest throughput among all schemes that utilize multiple paths due to absence of redundancy.
It should also be noted that network coding-based schemes that allow only one packet generation to be on the network each time achieve higher throughput than greedy network coding schemes.
This is because network under-utilization due to idle times between successive packet generations is balanced by the gain due to reduced interference.
NC-L for example, experiences $40.2$\% fewer dropped packets due to path loss and interference when compared to G-NC-L.

\begin{table}[h]
\begin{center}
\scriptsize
\begin{tabular}{|c|c|c|c|c|c|c|}
\hline
     & \multicolumn{2}{|c|}{Simulation}     & \multicolumn{4}{|c|}{Numerical}                                   \\ \hline
     & \multicolumn{2}{|c|}{$d_{h}=40m$} & \multicolumn{2}{|c|}{ Error=\{0.2, 0.4\}} & \multicolumn{2}{|c|}{ \{e1,e2,e3\}=\{$0.3$,$0.4$,$0.5$\} }\\ \hline
Rank & Delay               & Throughput              &  Delay           & Throughput  &  Delay           & Throughput          \\ \hline
1    & MC                 & MP                & MC              & MP   &  MC & MP      \\ \hline
2    & SP                 & G-NC                & NC              & NC & NC & NC  \\ \hline
3    & MP                & NC                & SP,MP              & MC &  SP & MC  \\ \hline
4    & NC                 & MC                &               & SP & MP & SP  \\ \hline
5    & G-NC                 & SP                &               &   &  &    \\ \hline
\end{tabular}
\end{center}
\caption{Numerical vs simulation results. Three single-hop paths}
\label{tab:comp_3}
\end{table}

Table \ref{tab:comp_3} contradicts the throughput and delay trends for the numerical results presented in tables \ref{table:trcommon}, \ref{table:trdiff} and the simulation results of table \ref{tab:sim_topol_4},
for the case of a network consisting of three single hop paths.

We first compare the simulation results with the numerical ones concerning the case of same link error probability for all links ($0.2$ or $0.4$) and hop-by-hop coding process
for network coding-based schemes.
We first observe that MC employing the maximum redundancy achieves the lowest delay.
The main difference between simulation and the numerical results concerns the rank of NC in terms of delay.
In the scenario simulated NC appears to experience higher delay than both SP and MP.
As also discussed in the simulation setup (section \ref{sec:sim_setup}), packets are injected into each path with a constant probability of $0.2$ independently of each other.
For a topology consisting of three single-hop paths and given that the transmission probability of each of the three interfaces assumed for the source node is $0.2$, the probability that two or more packets overlap is $10.4$\%.
Consequently it is not expected for packets to experience significant interference.
Indeed, in comparison with the scenario with three paths of four hops each discussed before in this section, the number of dropped frames due to path loss and interference is dramatically lower.
Moreover the scenario discussed considers single hop paths so packets do not experience any queuing delay.
The only overhead for each packet is the time spent at the source node waiting to be transmitted.
This overhead becomes more significant in the case of NC since it is aggregated over all the coded packets that are produced from a specific packet generation.
Replaying the same simulation scenario using a transmission probability of $0.3$ instead of $0.2$ for the source node, the delay of NC is reduced by 17\% approximately (the corresponding results can be found in the Appendix).
Another difference concerning delay between the numerical and the simulation results is that SP achieves lower delay than MP since it experiences significantly fewer dropped frames due to interference (shown in table \ref{tab:sim_topol_4}).
As far as throughput is concerned, our analytical framework accurately capture the throughput trend for all forwarding schemes.

As table \ref{tab:comp_3} also shows, When different link error probabilities are assumed, the suggested framework captures the throughput delay trade-off observed in the simulation results missing only the case of NC.
The reason for this, is the waiting times at the source node before transmission which are aggregated over all packets comprising a packet generation.
\begin{table}[h]
\begin{center}
\scriptsize
\begin{tabular}{|c|c|c|c|c|}
\hline
     & \multicolumn{2}{|c|}{Simulation}     & \multicolumn{2}{|c|}{Numerical}                                   \\ \hline
     & \multicolumn{2}{|c|}{$d_{h}=40m$,} & \multicolumn{2}{|c|}{Error=\{0.2, 0.4\}} \\ \hline
Rank & Delay               & Throughput              & Delay            & Throughput            \\ \hline
1    & SP                 & MP                & MC              & MP          \\ \hline
2    & MC                 & G-NC-L                & NC-L              & NC-L      \\ \hline
3    & MP                & NC-L                & NC-U              & NC-U   \\ \hline
4    & NC-L               & NC-U                & SP,MP              & MC   \\ \hline
5    & NC-U              & G-NC-U                &               & SP     \\ \hline
6    & G-NC-L            & SP               &               &           \\ \hline
7    & G-NC-U            & MC                &               &           \\ \hline
\end{tabular}
\end{center}
\caption{Numerical vs simulation results. Seven single-hop paths}
\label{tab:comp_4}
\end{table}

Table \ref{tab:comp_4} contradicts the throughput and delay trends for the numerical results presented in table \ref{table:trcommon} and the simulation results of table \ref{tab:sim_topol_5},
for the case of a network consisting of seven single hop paths.
As far as network coding based schemes are concerned hop-by-hop coding process is assumed in the analysis.

In the scenario where seven single hop paths are concerned instead of three, the probability of two or more packet transmissions overlapping increases and consequently transmitters experience increased interference.
This is also the reason for which SP achieves lower delay than MP in the simulation scenarios as opposed to the trend revealed by the numerical results.
More on the effect of interference on delay, simulation results reveal that MP achieves lower delay than NC-based schemes that allow only one packet generation into the network (NC-L, NC-U).
Indeed, in the scenario simulated MP experiences $19.1$\%fewer dropped packets due to path loss and interference than NC-L for example.
This is due to the significantly lower flow rate injected into each path by MP as opposed to network coding-based schemes.
As table \ref{tab:comp_4} shows, our analysis captures the trend in terms of throughput overestimating only MC's rank.
This is however expected since MC injects higher flow rates than all other multi-path schemes into each path imposing significant interference on the network.

\section{\uppercase{Conclusions}}
\label{sec:conclusions}

This paper presented an analytical framework for expressing the throughput and delay of various forwarding schemes employing multiple paths and different degrees of redundancy for wireless networks.
The analysis was first presented for a wireless erasure channel with link error probability being captured through the SINR model and demonstrated the complexity for generalizing for arbitrary topologies.
The analysis was also presented for a wireless erasure channel with link error probability being captured through the SNR model for different network settings depending on whether end-to-end coding is employed and node disjointness of paths.
The throughput and delay results captured by the analytical framework were validated and extended through NS-2 simulations.

Our results show that in scenarios where significant inter- and intra-path interference is present, the analytical framework presented capture the trends in terms of throughput and delay for network coding based
forwarding and multipath.
The most important deviation between the numerical and the simulation results concerns single path forwarding whose rank both in terms of delay and throughput is underestimated.
This is due to the SNR-based approximation of interference (instead of SINR-based) adopted in the analytical framework.
Multicopy also exhibits lower delay rank in the simulation scenarios due to increased interference.
More precisely, for the scenarios of three paths with four hops and seven paths with two hops the best throughput-delay trade-off is achieved by single path forwarding.
In the scenario with seven single hop paths where flow rate is distributed over more paths and thus interference is moderated, schemes that employ multiple paths and low redundancy are favored in terms of throughput.
Network coding experiences higher delay than all other schemes due to the inter-arrival times aggregated over all coded packets required to decode a packet generation.
Although interference is moderated by lower flow rate injected in each path still our analytical framework underestimates the rank only of single-path forwarding both in terms of delay and throughput.

In scenarios with lower interference our analytical framework underestimates only network coding rank in terms of delay.
Finally, flows of high data rate may lead to increased delay due to packets accumulating at the relay nodes.
This is more prominent in scenarios where fewer paths are employed to forward traffic to the destination as in the case of single-path forwarding.

\bibliographystyle{ieeetr}

\bibliography{bibliography}

\section*{\uppercase{Appendix}}
\label{sec:appendix}

\begin{table}[hb]
\begin{center}
\scriptsize
\begin{tabular}{|l|c|c|c|c|c|}
\hline
~                                    & MP & MC & NC & G-NC & SP \\ \hline
Delay (msec)   & 36.9  & 23.0  & 27.1  & 86.7 & 19.2  \\ \hline
Throughput                   & ~  & ~  & ~  & ~         & ~  \\
(Mbps)                   & 2.19  & 1.23  & 1.62 & 1.80 & 1.42 \\ \hline
Inter-arrival              & ~  & ~  & ~  & ~         & ~  \\
times (msec)   &   &   &  10.3 & 52.9 & \\ \hline
Pkt drops  & 8895  & 10175  & 5751  & 9089 & 1416 \\ \hline
\end{tabular}
\end{center}
\caption {Three paths, four hops. $d_{h}=40m$, $d_{v}=80m$. Transmission Probability = 0.1}
\label{tab:sim_topol_1_0.1}
\end{table}

\begin{table}[hb]
\begin{center}
\scriptsize
\begin{tabular}{|l|c|c|c|c|c|}
\hline
~                                    & MP & MC & NC & G-NC & SP \\ \hline
Delay (msec)   & 4111.8   & 1875.5 & 28.2 & 4881.3 & 1917.1 \\ \hline
Throughput                   & ~  & ~  & ~  & ~         & ~  \\
(Mbps)                   & 1.13  & 0.58  & 1.62 & 0.93 & 2.08 \\ \hline
Inter-arrival              & ~  & ~  & ~  & ~         & ~  \\
times (msec)   &   &   &  7.5 & 121.9 & \\ \hline
Pkt drops  & 59674  & 109012  & 17159  & 69494 & 4581 \\ \hline
\end{tabular}
\end{center}
\caption {Three paths, four hops. $d_{h}=40m$, $d_{v}=80m$. Transmission Probability = 0.3}
\label{tab:sim_topol_1_0.3}
\end{table}

\begin{table}[hb]
\begin{center}
\scriptsize
\begin{tabular}{|l|c|c|c|c|c|}
\hline
~                                    & MP & MC & NC & G-NC & SP \\ \hline
Delay (msec)   & 43.2 & 21.7 & 33.0 & 3810.5 & 19.2 \\ \hline
Throughput                   & ~  & ~  & ~  & ~         & ~  \\
(Mbps)                   & 1.97 & 1.21 & 1.34 & 1.30 & 1.42 \\ \hline
Inter-arrival              & ~  & ~  & ~  & ~         & ~  \\
times (msec)   &   &   &  15.8 & 2252.3 & \\ \hline
Pkt drops  & 9141  & 8728  & 7232  & 11387 & 1416 \\ \hline
\end{tabular}
\end{center}
\caption {Three paths, four hops. $d_{h}=40m$, $d_{v}=120m$. Transmission Probability = 0.1}
\label{tab:sim_topol_2_0.1}
\end{table}

\begin{table}[hb]
\begin{center}
\scriptsize
\begin{tabular}{|l|c|c|c|c|c|}
\hline
~                                    & MP & MC & NC & G-NC & SP \\ \hline
Delay (msec)   & 3478.0 & 4113.7 & 31.4 & 11089.7 & 1917.1 \\ \hline
Throughput                   & ~  & ~  & ~  & ~         & ~  \\
(Mbps)                   & 1.35  & 0.95 & 1.46 & 0.78 & 2.08 \\ \hline
Inter-arrival              & ~  & ~  & ~  & ~         & ~  \\
times (msec)   &   &   &  11.7 & 4145.7 & \\ \hline
Pkt drops  & 44664  & 49995  & 19239  & 76824 & 4581 \\ \hline
\end{tabular}
\end{center}
\caption {Three paths, four hops. $d_{h}=40m$, $d_{v}=120m$. Transmission Probability = 0.3}
\label{tab:sim_topol_2_0.3}
\end{table}

\begin{table}[hb]
\begin{center}
\scriptsize
\begin{tabular}{|l|c|c|c|c|c|c|c|}
\hline
~                                   & MP & MC & NC-L & NC-U & G-NC-L & G-NC-U & SP \\ \hline
Delay           &   &   &  &  & & & \\
(msec)           & 42.7  & 12.3  & 28.0 & 32.4 & 70.5 & 115.1 & 10.7\\ \hline
Throughput             & ~  & ~  & ~    & ~    & ~           & ~           & ~  \\
(Mbps)                 & 3.38  & 1.35 & 2.34 & 2.06 & 2.43 & 2.09 & 1.69  \\ \hline
Inter-arrival             & ~  & ~  & ~    & ~    & ~           & ~           & ~  \\
times (msec)   &   &   &  14.0 & 20.1 & 42.8 &  79.0 &   \\ \hline
Pkt drops          & 12689  & 16568  & 10201 & 10914 & 11596 & 15108 & 221 \\ \hline
\end{tabular}
\end{center}
\caption {Seven paths, two hops. $d_{h}=40m$, $d_{v}=10m$. Transmission Probability = 0.1}
\label{tab:sim_topol_3_0.1}
\end{table}

\begin{table}[hb]
\begin{center}
\scriptsize
\begin{tabular}{|l|c|c|c|c|c|c|c|}
\hline
~                                   & MP & MC & NC-L & NC-U & G-NC-L & G-NC-U & SP \\ \hline
Delay           &   &   &  &  & & & \\
(msec)           & 7400.0  & 199.1  & 84.8 & 82.5 & 9772.5 & 12815.6 & 1205.0\\ \hline
Throughput             & ~  & ~  & ~    & ~    & ~           & ~           & ~  \\
(Mbps)                 & 0.56  & 0.24 &  0.82   & 0.85  & 0.32  & 0.29 & 3.09  \\ \hline
Inter-arrival             & ~  & ~  & ~    & ~    & ~           & ~           & ~  \\
times (msec)   &   &   &  32.4 & 39.5 &  500.5 &  1225.6 &   \\ \hline
Pkt drops          & 216765  & 201671  & 107704 & 98541 & 370834 & 405743 & 792 \\ \hline

\end{tabular}
\end{center}
\caption {Seven paths, two hops. $d_{h}=40m$, $d_{v}=10m$. Transmission Probability = 0.3}
\label{tab:sim_topol_3_0.3}
\end{table}

\begin{table}[hb]
\begin{center}
\scriptsize
\begin{tabular}{|l|c|c|c|c|c|}
\hline
~                                    & MP & MC & NC & G-NC & SP \\ \hline
Delay           &   &   &  &  & \\
(msec)   & 8.0 & 3.7  & 11.5  & 22.0 & 6.4  \\ \hline
Throughput                   & ~  & ~  & ~  & ~         & ~  \\
(Mbps)                   & 4.42  & 2.38  & 3.11 & 3.47 & 1.84\\ \hline
Inter-arrival              & ~  & ~  & ~  & ~         & ~  \\
times (msec)   &   &   &  7.1 & 15.9 & \\ \hline
Pkt drops  & 662  & 979  & 623  & 725 & 35 \\ \hline
\end{tabular}
\end{center}
\caption {Three paths, one hop. $d_{h}=40m$. Transmission Probability = 0.1}
\label{tab:sim_topol_4_0.1}
\end{table}

\begin{table}[hb]
\begin{center}
\scriptsize
\begin{tabular}{|l|c|c|c|c|c|}
\hline
~                                    & MP & MC & NC & G-NC & SP \\ \hline
Delay           &   &   &  &  & \\
(msec)   & 4.2  & 1.9 & 5.8  & 13.6 & 2.0  \\ \hline
Throughput                   & ~  & ~  & ~  & ~         & ~  \\
(Mbps)                   & 8.29  & 4.55  & 6.16 & 6.58 & 5.67\\ \hline
Inter-arrival              & ~  & ~  & ~  & ~         & ~  \\
times (msec)   &   &   &  3.4 & 10.1 & \\ \hline
Pkt drops  & 2331  & 3839  & 2425  & 2767 & 34 \\ \hline
\end{tabular}
\end{center}
\caption {Three paths, one hop. $d_{h}=40m$. Transmission Probability = 0.3}
\label{tab:sim_topol_4_0.3}
\end{table}

\begin{table}[hb]
\begin{center}
\scriptsize
\begin{tabular}{|l|c|c|c|c|c|c|c|}
\hline
~                                   & MP & MC & NC-L & NC-U & G-NC-L & G-NC-U & SP \\ \hline
Delay           &  12.0 & 3.29 & 14.6 & 18.1 & 26.7 & 42.7 & 6.4\\
(msec)           &   &  &   &   &   &  &  \\ \hline
Throughput             & ~  & ~  & ~    & ~    & ~           & ~           & ~  \\
(Mbps)                 & 6.86 & 2.38  & 3.97 & 3.45 & 4.23 & 3.71 & 1.84 \\ \hline
Inter-arrival             & ~  & ~  & ~    & ~    & ~           & ~           & ~  \\
times (msec)   &   &   & 9.8 & 13.9 & 19.2 & 33.9 &  \\ \hline
Pkt drops       & 3349 & 5239 &  2817 & 2859  & 3032 & 3396 & 35 \\ \hline
\end{tabular}
\end{center}
\caption {Seven paths, one hop. $d_{h}=40m$. Transmission Probability = 0.1}
\label{tab:sim_topol_5_0.1}
\end{table}

\begin{table}[hb]
\begin{center}
\scriptsize
\begin{tabular}{|l|c|c|c|c|c|c|c|}
\hline
~                                   & MP & MC & NC-L & NC-U & G-NC-L & G-NC-U & SP \\ \hline
Delay           &  17.1 &  5.0 & 20.7 & 23.2 & 40.4 & 68.8 & 2.0\\
(msec)           &   &  &   &   &   &  &  \\ \hline
Throughput             & ~  & ~  & ~    & ~    & ~           & ~           & ~  \\
(Mbps)                 & 4.84 & 1.59 & 2.97 & 2.80 & 2.82 & 2.49 & 5.67\\ \hline
Inter-arrival             & ~  & ~  & ~    & ~    & ~           & ~           & ~  \\
times (msec)   &   &   & 12.4 & 16.0 & 27.5 & 54.3 &  \\ \hline
Pkt drops       & 16267 & 45494 &  22569 & 22431  & 25711 & 29151 & 34 \\ \hline
\end{tabular}
\end{center}
\caption {Seven paths, one hop. $d_{h}=40m$. Transmission Probability = 0.3}
\label{tab:sim_topol_5_0.3}
\end{table}

\end{document}